\documentclass[lettersize,journal]{IEEEtran}
\usepackage{amsmath,amsfonts}
\usepackage{cite}
\allowdisplaybreaks[4]
\usepackage[caption=false,font=footnotesize,labelfont=rm,textfont=rm]{subfig}
\usepackage{textcomp}
\usepackage{enumerate}
\usepackage{amssymb}
\usepackage{algorithm}
\usepackage{algpseudocode}
\usepackage{amsthm}
\usepackage{stfloats}
\usepackage{url}
\usepackage{verbatim}
\usepackage{graphicx}
\usepackage{bm}
\usepackage{array,color}
\usepackage{float}
\usepackage{multirow}
\usepackage{makecell}
\begin{document}
\hyphenpenalty=100
\newtheorem{proposition}{Proposition}
\newtheorem{lemma}{Lemma}
\title{Sensing Capacity for Integrated Sensing and Communication Systems in Low-Altitude Economy}
\author{
    Jiahua Wan, Hong Ren,~\IEEEmembership{Member,~IEEE,} Cunhua Pan,~\IEEEmembership{Senior Member,~IEEE,} Zhenkun Zhang, Songtao Gao, Yiming Yu, and Chengzhong Wang 
    \thanks{J. Wan, H. Ren, C. Pan, Z. Zhang, and C. Wang are with National Mobile Communications Research Laboratory, Southeast University, Nanjing 211111, China. (email:\{wanjiahua, hren, cpan, zhenkun\_zhang, 213221379\}@seu.edu.cn). S. Gao and Y. Yu are with China Mobile Group Design Institute Co., Ltd., Beijing 100080, China. (email:\{gaosongtao,yuyiming\}@cmdi.chinamobile.com).}
} 
\maketitle
\begin{abstract} 
    The burgeoning significance of the low-altitude economy (LAE) has garnered considerable interest, largely fuelled by the widespread deployment of unmanned aerial vehicles (UAVs). To tackle the challenges associated with the detection of unauthorized UAVs and the efficient scheduling of authorized UAVs, this letter introduces a novel performance metric, termed sensing capacity, for integrated sensing and communication (ISAC) systems. This metric, which quantifies the capability of a base station (BS) to detect multiple UAVs simultaneously, leverages signal-to-noise ratio (SNR) and probability of detection (PD) as key intermediate variables. Through mathematical derivations, we can derive a closed-form solution for the maximum number of UAVs that can be detected by the BS while adhering to a specific SNR constraint. Furthermore, an approximate solution based on PD constraints is proposed to facilitate the efficient determination of the threshold for the maximum number of detectable UAVs. The accuracy of this analytical approach is verified through extensive simulation results. 
\end{abstract}

%\vspace{-0.1cm}
\section{Introduction}
The low-altitude economy (LAE), characterized by a convergence of low-altitude flight operations utilizing both manned and unmanned aerial vehicles (UAVs), has garnered significant interest from both academia and industry in recent years due to its diverse applications in areas such as the Internet of Things (IoT), service provision, supply chain management, agricultural practices, and disaster surveillance \cite{LAE1,LAE2}. In practical scenarios, the increase in the number of UAVs presents a challenge in designing obstacle avoidance mechanisms, emphasizing the importance of establishing real-time communication, tracking, and coordination protocols for authorized UAVs within the LAE, along with robust surveillance measures to detect and address unauthorized UAV intrusions.

As a pivotal technology in the realm of the six-generation (6G) wireless networks, integrated sensing and communication (ISAC) has been regarded as a viable and efficient solution to bolster the functionality of the LAE \cite{LAE_ISAC}. Specifically, through the fusion of communication and radar sensing within a unified framework, ISAC can reuse the spectrum in both communications and radar frequencies, leading to great enhancements in spectral and energy efficiency as well as the reductions in hardware and signaling expenditures \cite{ISAC5,ISAC2,ISAC4}. The ISAC system integrates wireless connectivity with precise sensing capabilities to facilitate real-time communication and monitoring of authorized UAVs, as well as airspace surveillance for unauthorized intrusions.

In the context of the integration paradigm of ISAC, balancing communication and sensing performance, along with understanding theoretical performance limits, are crucial considerations in the advancement of ISAC technology \cite{ISAC3}. Specifically, a thorough examination of the underlying performance limitations can unveil potential discrepancies between existing ISAC technologies and ideal designs. Furthermore, illustrating the trade-offs in performance can offer valuable guidance and insights for protocol design and systematic analysis of practical ISAC systems. In the theoretical performance evaluation of ISAC systems, the selection of various performance metrics significantly influences the trade-offs involved. In the area of communications, spectral efficiency \cite{SE}, rate \cite{rate}, signal-to-interference-plus-noise ratio (SINR) \cite{SINR}, and capacity \cite{capacity} are commonly used as performance metrics. Channel capacity, in particular, which denotes the maximum achievable limit by a communication system, plays a significant role in performance analysis and transmission design of communication systems. Conversely, in the radar area, probability of detection (PD) \cite{PD}, Crámer-Rao Lower Bound (CRLB) \cite{CRLB}, beampattern \cite{BP}, and mutual information (MI) \cite{MI} are widely utilized as performance metrics. However, these performance metrics lack explicit representation of the monitoring and regulatory capabilities concerning UAVs within the ISAC system. Additionally, the absence of information on sensing capacity presents a challenge in evaluating the performance trade-offs within the ISAC system.

In this letter, we analyze a novel sensing performance metric termed sensing capacity. This concept was originally introduced in \cite{sensing_capability} as the maximum number of UAVs that a base station (BS) can detect concurrently. However, a precise characterization for its performance remains unexplored. Unlike conventional performance metrics, sensing capacity offers insights into the BS's capability in UAV detection, thereby providing a solid theoretical foundation for future investigations on performance trade-offs in ISAC systems. Specifically, to enable the identification and expulsion of unauthorized UAVs, we choose SNR and PD as intermediate parameters for sensing capacity. Considering the stochastic nature of the locations of UAVs, we utilize stochastic geometry theory to establish a quantitative connection between the number of UAVs within the BS's detection area and both SNR and PD. The relationship between SNR and sensing capacity is revealed with a closed-form solution, and a low-complexity binary search-based algorithm for determining the sensing capacity based on the PD threshold is proposed. Finally, simulation results are presented to evaluate the sensing capacity under different system settings.

\emph{Notation:} In this letter, scalars, vectors, and matrices are represented by lowercase italic letters, lowercase bold letters, and uppercase bold letters, respectively. $(\cdot)^{\mathrm{T}}$ and $(\cdot)^{\mathrm{H}}$ denote the transpose and Hermitian operators, respectively. The space of $N \times M$ complex matrices is denoted by $\mathbb{C}^{N \times M}$. A random vector that follows the circularly symmetric complex Gaussian distribution is symbolized as $\sim \mathcal{C} \mathcal{N} (\boldsymbol{\mu },\boldsymbol{\Sigma })$, where the symbol $\sim$ means ``distributed as", and $\boldsymbol{\mu }$ and $\boldsymbol{\Sigma }$ are the mean vector and the correlation matrix, respectively. Similarly, a random scalar obeying a real Gaussian distribution with mean value of $\mu $ and variance of $\sigma ^2$ is denoted as $\sim\mathcal{N}(\mu , \sigma ^2)$. $\mathrm{Tr}(\cdot)$ represents the trace of a matrix and $\mathbb{E} (\cdot)$ denotes statistical expectation. $\left\lVert \cdot \right\rVert$ and $\left\lvert \cdot \right\rvert $ represent the Euclidean norm of a vector and the absolute value of a scalar, respectively. $\otimes$ denotes the Kronecker product.
%\vspace{-1cm}
\section{System Model}
\begin{figure}[tb]
    \centering
    \subfloat[]{\includegraphics[width=2.8in]{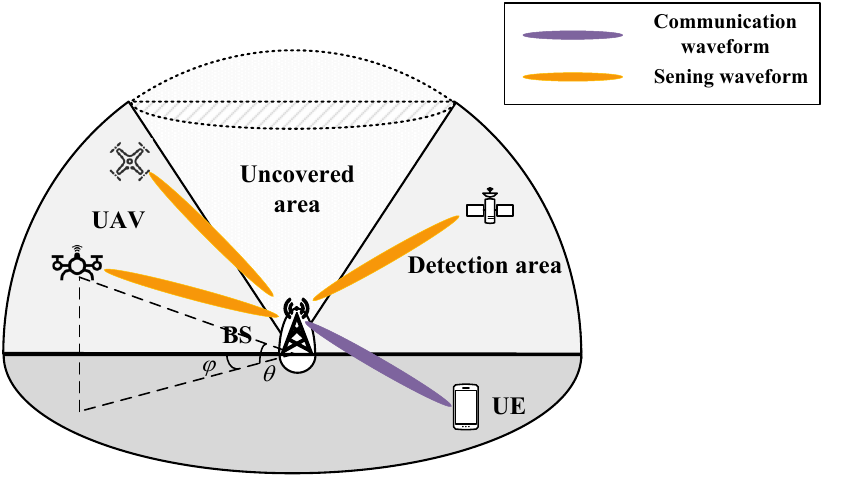} \label{fig1a}} 
    \\
    \subfloat[]{\includegraphics[width=2.6in]{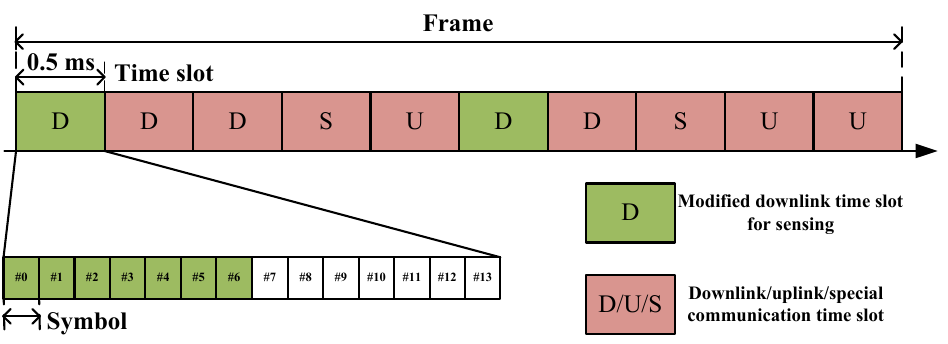} \label{fig1b}}
    \caption{(a) System model. (b) Modified frame structure of 5G NR.}
    \label{fig1}
\end{figure}

Consider an ISAC system, where an $M$-antenna BS utilizes the same frequency band for both radar and communication functionalities, aiming to detect $L$ UAVs within a modified frame derived from the existing 5G NR frame structure, as depicted in Fig. \ref{fig1}. Specifically, the designed frame consists of 10 time slots, with each time slot containing 14 symbols. Within the frame structure, the first and sixth time slots are designated as the modified downlink time slots for sensing, in which symbols from 0 to 6 are dedicated to sensing tasks. Thus, the total number of symbols allocated for sensing in a frame is 14. Meanwhile, the azimuth and elevation angles of the $l$-th UAV with respect to (w.r.t.) the BS are denoted as $\varphi_k,\theta_k,l=1,2,\cdots,L$, for all $l \in \mathcal{L} \triangleq \{1, \cdots, L\}$. The detection area of the BS is characterized by an elevation angle range $\theta_k \in [0, \Theta ]$ and the maximum distance $R$ at which the BS is capable of detecting the UAVs. Additionally, a small hemisphere with radius of $\frac{1}{\varepsilon}R$ is defined as the minimum distance constraint in this context\footnote{It is realistic since the reception processing of radar echoes involves a specific guard period.}. We assume that the UAVs\footnote{As the volume of the UAV is significantly smaller than the sensing area, each individual UAV is positioned at a considerable distance from one another, allowing for independent detection of each UAV.} are randomly and uniformly distributed within the hollow quarter sphere.
To mitigate the interference between uplink communication and sensing, the BS utilizes the time division duplex (TDD) method to perform both sensing and communication operations. Specifically, the BS simultaneously detects $K$ UAVs within a symbol\footnote{It is also known as pulse repetition interval in radar.} that are spatially well-separated \cite{ISAC3}, with no overlap of the main lobes in the beam direction towards each UAV. To improve the detection performance, $N$ consecutive symbols are utilized for continuously monitoring, referred to as a radar correlation processing interval (CPI). Hence, the total detection period of $L$ UAVs is equivalently expressed as $T = N \frac{L}{K}$ symbols. 

By defining the baseband signals as $\mathbf{s}_n(t) = [s_{1,n}(t), \cdots, s_{K,n}(t)]^{\mathrm{T}} \in \mathbb{C}^{K \times 1}$, satisfying $\mathbb{E} [\mathbf{s}_n \mathbf{s}_n ^{\mathrm{H}}] = {\mathbf{I}}_K$, the transmit signal at symbol $n$ is given by  
\begin{equation}
    \mathbf{x}_n(t) = \mathbf{F}_n \mathbf{s}_n(t) =\sum_{k=1}^{K} \mathbf{f}_{k,n} s_{k,n}(t) \in  \mathbb{C}^{M \times 1},
\end{equation}
where $\mathbf{F}_n = [\mathbf{f}_{1,n}, \cdots, \mathbf{f}_{K,n}] \in \mathbb{C}^{M \times K}$ denotes the precoding matrix. Decoding the radar echoes with a normalized received beamforming matirx $\mathbf{W}_n = [\mathbf{w}_{1,n}, \cdots, \mathbf{w}_{K,n}] ^{\mathrm{T}} \in \mathbb{C}^{M \times K}$, we can obtain the radar's received signal as 
\begin{equation}
    \label{equr}
    \begin{aligned}
        \mathbf{r}_n (t) &= \mathbf{W}_n^{\mathrm{H}} \sum_{k=1}^{K} \kappa \sqrt{P_T} \beta _{k} e^{j 2 \pi \omega_{k} t} \\
        & \times {\mathbf{a}_{\mathrm{r}}}(\varphi_k,\theta _k){\mathbf{a}_{\mathrm{t}}}^ {\mathrm{H}}(\varphi_k,\theta _k) {\mathbf{x}}(t-\tau _{k}) + \mathbf{W}_n^{\mathrm{H}} {\mathbf{z}}(t) \in  \mathbb{C}^{K \times 1},
    \end{aligned}
\end{equation} 
where $\kappa$ denotes the product of array gain and beamforming gain, $P_T$ represents the transmit power, and ${\mathbf{z}}(t) \sim \mathcal{C} \mathcal{N} (\bm{\mathrm{0}},\sigma^2\bm{\mathrm{I}}_M)$ represents the additive white Gaussian noise (AWGN). $\beta _{k}$, $\omega_{k}$, and $\tau _{k}$ denote the complex coefficient, the Doppler frequency and the time delay of UAV $k$, respectively. We assume that the $K$ UAVs within the same symbol have distinct range-Doppler $(\tau _{k}, \omega_{k})$ bins. ${\mathbf{a}_{\mathrm{r}}}(\varphi_k,\theta _k)$ and ${\mathbf{a}_{\mathrm{t}}}(\varphi_k,\theta _k)$ represent the receive and transmit steering vectors, respectively. Since we focus on LAE scenarios, we define ${\mathbf{H}_{k}} \triangleq \beta_k {\mathbf{a}_{\mathrm{r}}}(\varphi_k,\theta _k){\mathbf{a}_{\mathrm{t}}}^ {\mathrm{H}}(\varphi_k,\theta _k)\in \mathbb{C}^{M \times M}$.

Based on the characteristic of the ISAC-enabled BS, we have ${\mathbf{a}_{\mathrm{r}}}(\varphi_k,\theta _k) = {\mathbf{a}_{\mathrm{t}}}(\varphi_k,\theta _k) \triangleq {\mathbf{a}}(\varphi_k,\theta _k)$. Moreover, without loss of generality, it is assumed that the BS is equipped with a uniform planar array (UPA) with size of $M=M_x \times M_y$, where $M_x$ and $M_y$ represent the number of antennas along the x-axis and y-axis, respectively. Hence, the array response vector can be expressed as
\begin{equation}
    {\mathbf{a}}(\varphi_k,\theta _k) = {\mathbf{a}}^{(e)}(\varphi_k,\theta _k) \otimes {\mathbf{a}}^{(a)}(\varphi_k,\theta _k),
\end{equation}
where $\otimes$ denotes the Kronecker product, and ${\mathbf{a}}^{(e)}(\varphi_k,\theta _k)$ and ${\mathbf{a}}^{(a)}(\varphi_k,\theta _k)$ are respectively given as follows
\begin{align}
    &{\mathbf{a}}^{(e)}(\varphi_k,\theta _k) = \sqrt{\frac{1}{M_y}}\left[ 1, \cdots, e ^{\frac{j 2 \pi d (M_y-1) \sin (\varphi _k)\sin (\theta _k) }{\lambda}} \right] ^T,\\
    &{\mathbf{a}}^{(a)}(\varphi_k,\theta _k) = \sqrt{\frac{1}{M_x}}\left[ 1, \cdots, e ^{\frac{j 2 \pi d (M_x-1) \sin (\varphi _k) \cos (\theta _k)}{\lambda}} \right] ^T,
\end{align}
where $d$ and $\lambda$ represent the distance between the adjacent elements of the BS and the carrier wavelength, respectively. For simplicity, we set $d = \lambda/2$ in this letter.

The path loss $\beta_k$ can be modeled according to the large-scale channel path loss in free space \cite{ITU-R}
\begin{equation}
    \beta_k [\textrm{dB}] = 103.4 + 20 \log f + 40 \log d_k - 10 \log \sigma _{RCS},
\end{equation}
where $f$ represents the system operation frequency of the BS (in MHz), $d_k$ represents the distance between the BS and UAV $k$ (in km), and $\sigma _{RCS}$ denotes the radar cross section (RCS) of the UAV (in $\mathrm{m^2}$), which is set as the same for all UAVs for simplicity.

Then, for the decoding vector $\mathbf{w}_{k}$ and the precoding vector $\mathbf{f}_{k}$, the maximum ratio combining (MRC) criterion is applied to derive the optimal solution as 
\begin{equation}
    \mathbf{w}_{k} = \frac{{\mathbf{a}_{\mathrm{r}}}(\varphi_k,\theta _k)}{\left\lVert {\mathbf{a}_{\mathrm{r}}}(\varphi_k,\theta _k)\right\rVert _2}, \mathbf{f}_{k} = \frac{{\mathbf{a}_{\mathrm{t}}}(\varphi_k,\theta _k)}{\sqrt{K}\left\lVert {\mathbf{a}_{\mathrm{t}}}(\varphi_k,\theta _k)\right\rVert _2}.
\end{equation}

Since each UAV is uniquely identified by a specific range-Doppler bin, employing spatial-temporal matched filtering \cite{matched_filter} on the received signal in    (\ref{equr})  within a designated range-Doppler bin $(\tau _{k}, \omega_{k})$ enables the extraction of UAV $k$'s signal
\begin{equation}
    y_{k,n} = \kappa \sqrt{\frac{P_T}{K}} \beta _k + z_{k,n},
\end{equation}
where $y_{k,n} = \int r_{k,n} (t) s_{k,n} ^*(t-\tau _k) e^{-j 2 \pi \omega_{k} t} dt$, $r_{k,n} (t)$ denotes the $k$-th element of $\mathbf{r}_n (t)$. $z_{k,n} = \int \mathbf{w}_k^{\mathrm{H}} {\mathbf{z}}(t) s_{k,n} ^*(t-\tau _k) e^{-j 2 \pi \omega_{k} t} dt$ is the match filtering noise, satisfying $z_{k,n} \sim \mathcal{C}\mathcal{N} (0,\sigma^2)$. 
Given that ${\mathbf{H}_{k}}$ does not change over one CPI, we improve signal quality by overlaying the received echoes related to UAV $k$ during the $N$ symbols for signal detection. Then, by applying the coherent integration technique\cite{coherent_integration}, the cumulative energy of the received signal within $N$ symbols can be formulated as
\begin{equation}
    \begin{aligned}
        &\mathbb{E} \left[\left( \sum_{n=1}^{N} y_{k,n}\right)^* \left(\sum_{n=1}^{N} y_{k,n} \right)\right]  = \underbrace{\kappa ^2 \frac{P_T}{K} N^2 \beta _k^2}_{\mathrm{radar \, signal}} + \underbrace{N \sigma^2} _{\mathrm{noise}}.
    \end{aligned}
\end{equation}
Accordingly, the SNR of the received signal from UAV $l$ over the duration of $N$ symbols is given by 
\begin{equation}
    \begin{aligned}
        \label{equsnr}
        \mathrm{SNR}_l  = \frac{\kappa ^2 \frac{P_T}{K} N^2 \beta _l^2}{N \sigma^2} = \frac{\kappa ^2 N P_T}{K \epsilon  d_l^{4 } \sigma^2},
    \end{aligned}
\end{equation}
where $\epsilon \triangleq \frac{10^{10.34} f^2}{\sigma _{RCS}}$. As indicated in (\ref{equsnr}), the SNR of UAV $l$ depends on its distance from the BS. Given the assumption that UAV $l$ is randomly and uniformly distributed within the detection area, the distance $d_l$ is a random value. Thus, we will derive the average SNR of UAV $l$ based on its statistical properties within this area. By utilizing the probability density function (PDF) for UAV positioning, as outlined in (\ref{equ4}) at the top of the next page, the average SNR of UAV $l$ can be calculated as
\begin{figure*}
    \begin{equation}
        \label{equ4}
        f_d(x,\theta ,\varphi) = \left\{
            \begin{aligned}
                &\frac{3 \varepsilon^3 x^2 \cos\theta }{\pi  R^3(\varepsilon^3 -1)},\frac{1}{\varepsilon}R \leq x \leq R, 0 \leq \theta \leq \Theta , 0 \leq \varphi  \leq \pi \\
                &0,else.
            \end{aligned}
            \right.       
    \end{equation}
    \hrulefill
\end{figure*} 
\begin{equation}
    \label{equ6}
    \begin{aligned}
        \mathrm{SNR}_l &= \int_{0}^{\pi} \int_{0}^{\Theta } \int_{\frac{1}{\varepsilon}R}^R \frac{1}{ \delta d_l^4} f_d(d_l,\theta ,\varphi) dd_l d \theta d \varphi  \\
        & = \frac{1}{\delta } \int_{0}^{\pi} \int_{0}^{\Theta } \int_{\frac{1}{\varepsilon}R}^R \frac{3 \varepsilon^3 d_l^2 \cos\theta }{\pi  R^3(\varepsilon^3 -1)}   \frac{1}{d_l^4}dd_l d \theta d \varphi \\
        & = \frac{3 \varepsilon^3 \kappa ^2 N P_T  \sin\Theta }{K R^4 (\varepsilon^2+\varepsilon+1) \epsilon \sigma^2 },
    \end{aligned}
\end{equation} 
where $\frac{1}{\delta } \triangleq  \frac{\kappa ^2 N P_T}{K\epsilon \sigma^2}$. Since the UAVs are uniformly distributed in the detection area, the average SNR for each UAV remains consistent. By using the equation $T = N\frac{L}{K}$, the average SNR for $L$ UAVs detected by the BS during period $T$, denoted as $\mathrm{SNR}_L$, can be derived as 
\begin{equation}
    \label{equsnrl}
    \mathrm{SNR}_L = \frac{3 \varepsilon^3 \kappa ^2 T P_T  \sin\Theta }{L R^4 (\varepsilon^2+\varepsilon+1) \epsilon \sigma^2 }.
\end{equation}
\textit{Remark: The above result establishes an relationship between the number of detected UAVs, i.e., the sensing capacity, and the average SNR. The statistical features reveal more design insights into the factors affecting sensing capacity  compared to numerical results derived from Monte Carlo simulation. It can be seen that the detection radius $R$ of the BS is a predominant factor influencing sensing capacity.}
%\vspace{-1cm}
\section{Probability of Detection}
This letter focuses on identifying the maximum number of UAVs that can be detected by the BS, while satisfying the sensing requirements. To this end, PD is used as a key performance indicator, which will be derived in this section.

For UAV $l$, the detection problem can be defined as a binary hypothesis testing problem. By considering the null hypothesis $\mathcal{H} _0$ and alternative hypothesis $\mathcal{H} _1$, the detection model for the output signal $y_l$ can be expressed as 
\begin{equation}
    y_l = \left\{
        \begin{aligned}
            &\mathcal{H} _0:  z_l \\
            &\mathcal{H} _1:  \kappa \sqrt{\frac{P_T}{K}} \beta _l + z_l.
        \end{aligned}
        \right.       
\end{equation}
The Neyman-Pearson lemma states that the likelihood ratio test (LRT) maximizes the PD while maintaining a constant probability of false alarm (PFA) \cite{PD}. Specifically, the logarithmic LRT function is formulated as 
\begin{equation}
    \begin{aligned}
        \ln \Lambda (y_l) 
        &= \frac{1}{\sigma^2}\left(2 \mathrm{Re} \left( y_l^* \kappa \sqrt{\frac{P_T}{K}} \beta _l \right)  - \kappa^2 \frac{P_T}{K}\beta _l^2\right). \\
    \end{aligned}
\end{equation}
The distribution of $\mathrm{Re}\left(y_l^* \kappa \sqrt{\frac{P_T}{K}} \beta _l\right) $ under hypotheses $\mathcal{H}_0$ and $\mathcal{H}_1$ are respectively expressed as
    \begin{equation} 
        \left\{
            \begin{aligned}
                & \mathrm{Re} \left(y_l^* \kappa \sqrt{\frac{P_T}{K}} \beta _l\right) \sim \mathcal{N} \left(0, \frac{1}{2} \sigma^2 \kappa^2 \frac{P_T}{K} \beta _l^2 \right) , \quad \mathcal{H} _0 \\
                & \mathrm{Re} \left(y_l^* \kappa \sqrt{\frac{P_T}{K}} \beta _l\right) \sim \mathcal{N}\left(\kappa^2 \frac{P_T}{K} \beta _l^2,\frac{1}{2} \sigma^2 \kappa^2 \frac{P_T}{K} \beta _l^2\right) , \quad \mathcal{H} _1.
            \end{aligned}
            \right.   
    \end{equation}
By defining the decision threshold as $\gamma$, the PFA can be obtained as 
\begin{equation}
    \begin{aligned}
        \mathrm{Pr}_{{FA},l} &= \mathrm{Pr}(\ln \Lambda (y_l) >\gamma |\mathcal{H} _0) \\
        &=  Q\left(\frac{\gamma}{\sqrt{2 \mathrm{SNR}_l}}+\sqrt{\frac{\mathrm{SNR}_l}{2}} \right), 
    \end{aligned}
\end{equation}
where $Q(x) = \frac{1}{\sqrt{2 \pi}} \int_x^{\infty } e^{-\frac{u^2}{2}} du$ denotes the tail distribution function for the standard normal distribution. Therefore, the detection threshold at a given $\mathrm{Pr}_{{FA},l}$ is expressed as 
\begin{equation}
    \gamma = \sqrt{2 \mathrm{SNR}_l}Q^{-1}(\mathrm{Pr}_{{FA},l})-\mathrm{SNR}_l.
\end{equation}
Then, by defining $\xi = Q^{-1}(\mathrm{Pr}_{FA})$, the detection probability is derived as
\begin{equation}
    \begin{aligned}
        \mathrm{Pr}_{{D},l} &= \mathrm{Pr}(\ln \Lambda (y_l) >\gamma |\mathcal{H} _1) = Q\left(\xi-\sqrt{2 \mathrm{SNR}_l} \right),
    \end{aligned}
\end{equation}
which is determined by the SNR of UAV $l$. Considering the uniform random distribution of each UAV, which results in an identical average SNR for all UAVs, as illustrated in (\ref{equsnrl}), the likelihood of successful detection of all UAVs by the BS is given by 
\begin{equation} 
    \label{equpd}
    \prod _{l=1}^L \mathrm{Pr}_{{D},l} = \left(Q\left(\xi-\sqrt{2 \mathrm{SNR}_L} \right)\right) ^L.
\end{equation}
By defining the threshold requirement for PD as $\varpi$, an equation can be established between $\varpi$ and the number of UAVs, yielding
\begin{equation}
    \label{equ16}
    L \ln Q\left(\xi-\sqrt{\frac{\varrho}{L} }\right)= \ln \varpi,
\end{equation}
where $\varrho \triangleq  \frac{6 \varepsilon^3 \kappa ^2 T P_T  \sin\Theta }{R^4 (\varepsilon^2+\varepsilon+1) \epsilon \sigma^2 } $. Next, we aim to utilize the approximation functions to simplify (\ref{equ16}). Since the PD thresholds typically range from 0.5 to 1 in practice, the corresponding values of the Q-function are taken within the interval $-4 \leq x \leq 0$. According to \cite{qfunc_appro}, the Q-function can be approximated by a second-order exponential function for $0 \leq x \leq 4$ as $Q(x) \approx e^{-ax^2-bx-c}$, where $a = 0.3842, b = 0.7640, c = 0.6964$. %, as shown in Fig. \ref{fig4}. 
Therefore, for $-4 \leq x \leq 0$, we have
\begin{equation}
    Q(x) = 1-Q(-x) \approx 1-e^{-0.3842x^2+0.7640x-0.6964},
\end{equation}
where the first equation is obtained by the property $Q(-x)+Q(x)=1$. Since $\ln(1-e^x) \approx -e^x$, the left-hand side of the equation in  (\ref{equ16}) can be approximated as
\begin{equation}
    f(L) = -L e^{-0.3842\frac{\varrho}{L}+(0.7684\xi-0.764)\sqrt{\frac{\varrho}{L}}+0.3798\xi-0.6964}. 
\end{equation}
\iffalse
\begin{figure}[h]
    \centering
    \includegraphics[width=2.6in]{Q.png}
    \caption{The Q-function and its approximation.}
    \label{fig4}
\end{figure}
\fi
It can be verified that the first-order derivative of $f(L)$ satisfies $f'(L) \leq 0$. Hence, $f(L)$ is monotonically decreasing. The binary search algorithm can be used to find the number of UAVs that meet the conditions, and the pseudocode for the algorithm is shown in Algorithm \ref{alg:algorithm1}.
\begin{algorithm}[t]
    \caption{Binary Search Algorithm}
    \label{alg:algorithm1}
    \begin{algorithmic}[1]
     %\fontsize{7pt}{8pt}
      \State Initialize the search boundaries $[L_{l}, L_{h}]$.
      \State  \textbf{Repeat} 
      \State \quad if $f\left(\left\lfloor \frac{L_{l}+L_{h}}{2} \right\rfloor \right) \leq \ln \varpi$
      \State \quad \quad $L_{h} \leftarrow \left\lfloor \frac{L_{l}+L_{h}}{2} \right\rfloor $
      \State \quad else
      \State \quad \quad $L_{l} \leftarrow \left\lfloor \frac{L_{l}+L_{h}}{2} \right\rfloor $
      \State \textbf{Until} $f\left(\left\lfloor \frac{L_{l}+L_{h}}{2} \right\rfloor \right) \geqslant \ln \varpi$ and $f\left(\left\lfloor \frac{L_{l}+L_{h}}{2} \right\rfloor + 1\right) \leq \ln \varpi$
    \end{algorithmic}
\end{algorithm}
\section{Simulation Results}
Numerical results are presented to evaluate the sensing capacity of the ISAC system. Simulation parameters are set as follows: $N = 3$, $M = 384$, $P_T = 58$ dBm, $\sigma _{RCS} = 0.01 \, \mathrm{m^2}$, $\kappa = 22.5$ dB. The system carrier frequency is set to $f = 4.9$ GHz. The default detection radius $R$ of the BS is set to 1 km, while the elevation angle limit of $\Theta $ is set to $\pi/5$. The system bandwidth is set to 100 MHz, and the noise power density is specified as -174 dBm/Hz. Unless otherwise stated, the standard time interval utilized for calculating the sensing capacity is set to 1 frame. The detection thresholds for SNR, PFA and PD are set to 13 dB, 0.05 and 0.95, respectively. 

\begin{figure}[h]
    \centering
    \subfloat[]{\includegraphics[width=4.3cm]{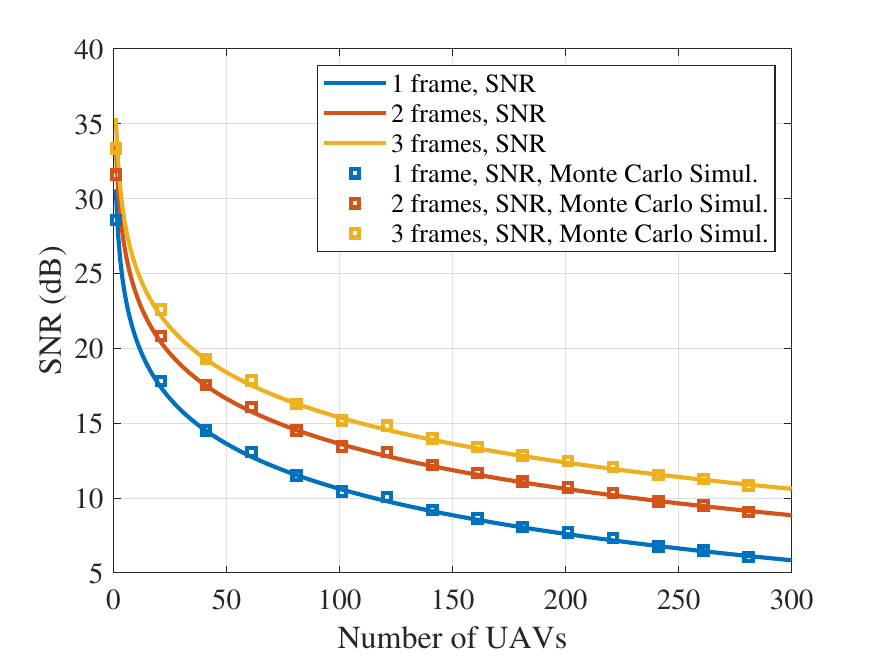} \label{snr}}
    \subfloat[]{\includegraphics[width=4.3cm]{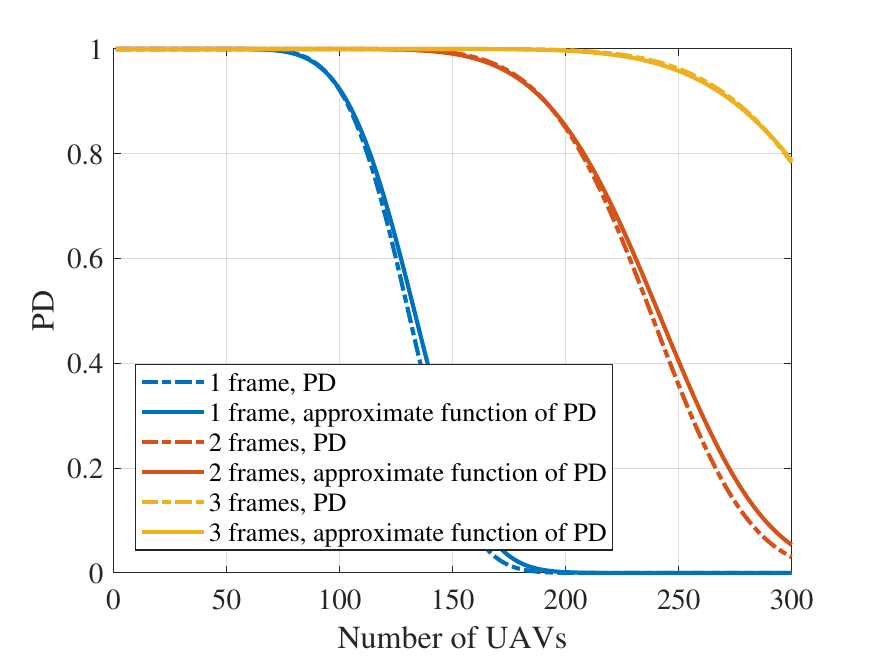}\label{pd}} 
    \caption{(a) SNR and (b) PD versus the number of UAVs.}
    \label{figframe}
\end{figure}

To investigate the impact of different SNR and PD thresholds on sensing capacity, Fig. \ref{figframe} illustrates the relationship between both the SNR and PD and the number of UAVs detected by the BS under varying numbers of frames. First, we verify the correctness of (\ref{equsnrl}) through Monte Carlo simulation, as depicted in Fig. \ref{figframe}-\subref{snr}. Based on the PDF of UAVs, $l$ UAVs are randomly generated within the detection area, and the average SNR detected by the BS is calculated. By averaging over $10^5$ generations, the SNR of the BS detecting $l$ UAVs is obtained. It can be observed that the Monte Carlo simulation results almost align with the theoretical values. Meanwhile, Fig. \ref{figframe}-\subref{pd} demonstrates that when the PD threshold is set at a high level, the results obtained from the approximation function closely match the exact values of PD derived from (\ref{equpd}), confirming the accuracy of our approximation method. Secondly, for a specific detection frame, the SNR decreases as the number of detected UAVs increases, and decreases slowly when the number of detected UAVs is large. This trend results from the inverse numerical relationship between SNR and the number of UAVs, as described in (\ref{equsnrl}). However, the PD initially declines slowly, followed by a sharp decrease, w.r.t. the number of UAVs, implying that there is a critical detection threshold, beyond which the detection performance is not acceptable. Finally, as expected, the number of detected UAVs exhibits an almost linear increase w.r.t. the number of frames for both specified PD and SNR thresholds.

\begin{figure}[h]
    \centering
    \includegraphics[width=3in]{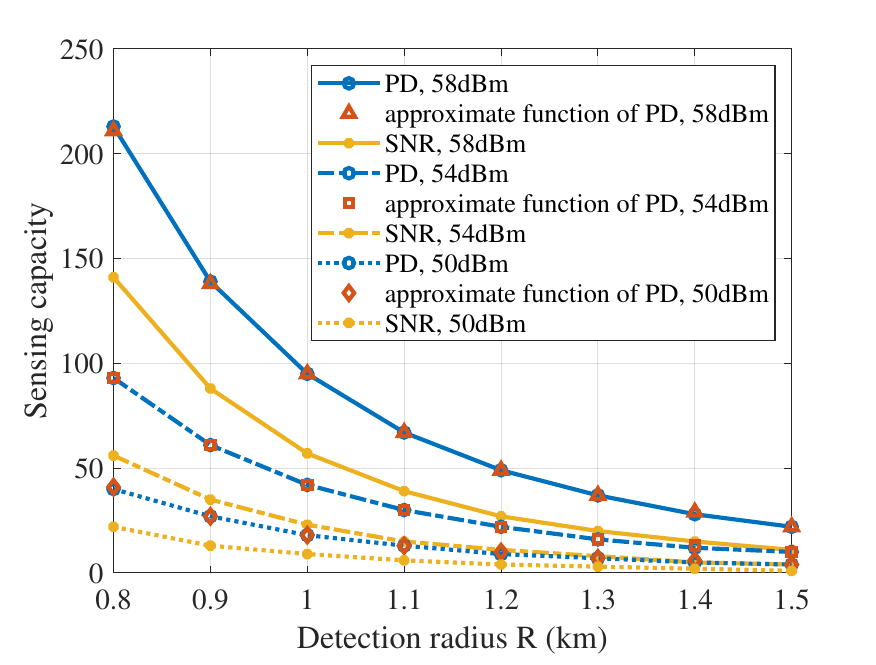}
    \caption{Sensing capacity versus the radius of the detection area constrained by PD and SNR.}
    \label{figd}
\end{figure}

%This observation underscores the pivotal role of the detection radius of the BS as a limiting factor in sensing capacity, highlighting the necessity for careful evaluation of the BS's detection radius to enhance sensing capacity in practical applications.
Fig. \ref{figd} depicts the sensing capacity versus the detection radius under scenarios with different transmit power. It can be seen that sensing capacity, measured by both PD and SNR, diminishes as the detection radius increases. This is because the volume of the detection area increases cubically with the radius, resulting in an increased challenge for the BS in UAV detection. Notably, when the transmit power of the BS is 50 dBm, the sensing capacity shows only a slight decline with the detection radius. This minor decrease occurs because the BS requires more symbols to detect the UAV while meeting the detection threshold at lower transmit power. Since the gain brought by increasing symbols exceeds the performance degradation caused by an enlarged detection radius, the dominant factor affecting the sensing capacity shifts to the length of the symbols, i.e., the detection frame. Furthermore, a reduction in sensing capacity under both constraints is observed with a decrease in transmit power. 
\section{Conclusion}
In this letter, we investigated a novel sensing performance metric, termed sensing capacity, to address the dual challenges of detecting unauthorized UAVs and optimizing the scheduling of authorized UAVs in ISAC systems. By highlighting the capability of the BS to  detect multiple UAVs concurrently, we utilized SNR and PD as key intermediate parameters to quantify the sensing capacity. Specifically, we derived a closed-form solution for SNR and the number of detected UAVs, and proposed an approximate solution for PD to efficiently determine the UAV threshold using a binary search algorithm. The effectiveness of our approach was validated through numerical results.
\bibliographystyle{IEEEtran}
\bibliography{IEEEabrv,Sensing_Capacity.bib}

\begin{thebibliography}{10}
\providecommand{\url}[1]{#1}
\csname url@samestyle\endcsname
\providecommand{\newblock}{\relax}
\providecommand{\bibinfo}[2]{#2}
\providecommand{\BIBentrySTDinterwordspacing}{\spaceskip=0pt\relax}
\providecommand{\BIBentryALTinterwordstretchfactor}{4}
\providecommand{\BIBentryALTinterwordspacing}{\spaceskip=\fontdimen2\font plus
\BIBentryALTinterwordstretchfactor\fontdimen3\font minus
  \fontdimen4\font\relax}
\providecommand{\BIBforeignlanguage}[2]{{%
\expandafter\ifx\csname l@#1\endcsname\relax
\typeout{** WARNING: IEEEtran.bst: No hyphenation pattern has been}%
\typeout{** loaded for the language `#1'. Using the pattern for}%
\typeout{** the default language instead.}%
\else
\language=\csname l@#1\endcsname
\fi
#2}}
\providecommand{\BIBdecl}{\relax}
\BIBdecl

\bibitem{LAE1}
C.~Xu, X.~Liao, J.~Tan, H.~Ye, and H.~Lu, ``Recent research progress of
  unmanned aerial vehicle regulation policies and technologies in urban low
  altitude,'' \emph{IEEE Access}, vol.~8, pp. 74\,175--74\,194, Apr. 2020.

\bibitem{LAE2}
N.~Hossein~Motlagh, T.~Taleb, and O.~Arouk, ``Low-altitude unmanned aerial
  vehicles-based internet of things services: Comprehensive survey and future
  perspectives,'' \emph{IEEE Internet Things J.}, vol.~3, no.~6, pp. 899--922,
  Dec. 2016.

\bibitem{LAE_ISAC}
G.~Cheng, X.~Song, Z.~Lyu, and J.~Xu, ``Networked {ISAC} for low-altitude
  economy: Transmit beamforming and {UAV} trajectory design,'' in \emph{2024
  IEEE/CIC International Conference on Communications in China (ICCC)}, Aug.
  2024, pp. 78--83.

\bibitem{ISAC5}
Z.~Zhang, H.~Ren, C.~Pan, S.~Hong, D.~Wang, J.~Wang, and X.~You, ``Target
  localization in cooperative {ISAC} systems: A scheme based on {5G} {NR}
  {OFDM} signals,'' 2024, [Online]. Available:
  https://arxiv.org/abs/2403.02028v2.

\bibitem{ISAC2}
Y.~Zhang, H.~Ren, C.~Pan, B.~Wang, Z.~Yu, R.~Weng, T.~Wu, and Y.~He, ``Secure
  wireless communication in active {RIS}-assisted {DFRC} systems,'' \emph{IEEE
  Trans. Veh. Technol.}, pp. 1--15, Aug. 2024.

\bibitem{ISAC4}
Z.~Yu, H.~Ren, C.~Pan, G.~Zhou, B.~Wang, M.~Dong, and J.~Wang, ``Active
  {RIS}-aided {ISAC} systems: Beamforming design and performance analysis,''
  \emph{IEEE Trans. Commun.}, vol.~72, no.~3, pp. 1578--1595, Mar. 2024.

\bibitem{ISAC3}
F.~Dong, F.~Liu, Y.~Cui, W.~Wang, K.~Han, and Z.~Wang, ``Sensing as a service
  in {6G} perceptive networks: A unified framework for {ISAC} resource
  allocation,'' \emph{IEEE Trans. Wireless Commun.}, vol.~22, no.~5, pp.
  3522--3536, May 2023.

\bibitem{SE}
Z.~Xiao and Y.~Zeng, ``Waveform design and performance analysis for full-duplex
  integrated sensing and communication,'' \emph{IEEE J. Select. Areas Commun.},
  vol.~40, no.~6, pp. 1823--1837, Jun. 2022.

\bibitem{rate}
F.~Liu, Y.~Cui, C.~Masouros, J.~Xu, T.~X. Han, Y.~C. Eldar, and S.~Buzzi,
  ``Integrated sensing and communications: Toward dual-functional wireless
  networks for {6G} and beyond,'' \emph{IEEE J. Select. Areas Commun.},
  vol.~40, no.~6, pp. 1728--1767, Jun 2022.

\bibitem{SINR}
B.~Zinat, T.~D. Özlem, W.~S. Ki, B.~Emil, and C.~Cicek, ``Multi-static target
  detection and power allocation for integrated sensing and communication in
  cell-free massive {MIMO},'' 2024, [Online]. Available:
  https://arxiv.org/abs/2305.12523.

\bibitem{capacity}
A.~Liu, Z.~Huang, M.~Li, Y.~Wan, W.~Li, T.~X. Han, C.~Liu, R.~Du, D.~K.~P. Tan,
  J.~Lu, Y.~Shen, F.~Colone, and K.~Chetty, ``A survey on fundamental limits of
  integrated sensing and communication,'' \emph{IEEE Commun. Surveys Tuts.},
  vol.~24, no.~2, pp. 994--1034, Feb. 2022.

\bibitem{PD}
\BIBentryALTinterwordspacing
S.~Kay, \emph{Fundamentals of Statistical Signal Processing: Detection theory},
  ser. Fundamentals of Statistical Si.\hskip 1em plus 0.5em minus 0.4em\relax
  Prentice-Hall PTR, 1998. [Online]. Available:
  \url{https://books.google.co.jp/books?id=vA9LAQAAIAAJ}
\BIBentrySTDinterwordspacing

\bibitem{CRLB}
M.~Tian, Y.~Liu, C.~Li, W.~Peng, T.~Wu, C.~Duan, and C.~Chen, ``Multiobjective
  optimal waveform design for {TDS-OFDM} integrated radar and communication
  systems,'' in \emph{2021 CIE International Conference on Radar (Radar)}, Dec.
  2021, pp. 2907--2911.

\bibitem{BP}
Z.~Liu, S.~Aditya, H.~Li, and B.~Clerckx, ``Joint transmit and receive
  beamforming design in full-duplex integrated sensing and communications,''
  \emph{IEEE J. Select. Areas Commun.}, vol.~41, no.~9, pp. 2907--2919, Sep.
  2023.

\bibitem{MI}
X.~Yuan, Z.~Feng, J.~A. Zhang, W.~Ni, R.~P. Liu, Z.~Wei, and C.~Xu,
  ``Spatio-temporal power optimization for {MIMO} joint communication and radio
  sensing systems with training overhead,'' \emph{IEEE Trans. Veh. Technol.},
  vol.~70, no.~1, pp. 514--528, Jan. 2021.

\bibitem{sensing_capability}
G.~Liu, L.~Ma, Y.~Xue, L.~Han, R.~Xi, Z.~Han, H.~Wang, J.~Dong, M.~Lou, J.~Jin,
  Q.~Wang, and Y.~Yuan, ``{SensCAP}: A systematic sensing capability
  performance metric for {6G} {ISAC},'' \emph{IEEE Internet Things J.},
  vol.~11, no.~18, pp. 29\,438--29\,454, Sep. 2024.

\bibitem{ITU-R}
ITU-R, \emph{Recommendation {ITU-R P}.525-2: Calculation of free-space
  attenuation}.\hskip 1em plus 0.5em minus 0.4em\relax Geneva: International
  Telecommunication Union, 2007.

\bibitem{matched_filter}
E.~Fishler, A.~Haimovich, R.~Blum, L.~Cimini, D.~Chizhik, and R.~Valenzuela,
  ``Spatial diversity in radars—models and detection performance,''
  \emph{IEEE Trans. Signal Process.}, vol.~54, no.~3, pp. 823--838, Mar. 2006.

\bibitem{coherent_integration}
M.~Richards, ``Coherent integration loss due to white gaussian phase noise,''
  \emph{IEEE Signal Process. Lett.}, vol.~10, no.~7, pp. 208--210, Jul. 2003.

\bibitem{qfunc_appro}
M.~López-Benítez and F.~Casadevall, ``Versatile, accurate, and analytically
  tractable approximation for the gaussian {Q}-function,'' \emph{IEEE Trans.
  Commun.}, vol.~59, no.~4, pp. 917--922, Apr. 2011.

\end{thebibliography}
\end{document}